\documentclass[review]{elsarticle}

\journal{Nuclear Instruments \& Methods in Physics Research, Section A}

\newcommand{\Gfi}{\ensuremath{G_{\mbox{\textit{fi}}}}}
\newcommand{\Co}{\ensuremath{^{60}}Co}
\usepackage[load-configurations = abbreviations]{siunitx}
\usepackage{xcolor}
\usepackage{subcaption}
\usepackage{amsmath}

\begin{document}

\begin{frontmatter}

\title{Ionization parameters of Trimethylbismuth for high-energy photon detection}





\author[mainaddress]{M. Farrad{\`e}che\corref{correspondingauthor}}\ead{morgane.farradeche@cea.fr}
\author[mainaddress]{G.~Tauzin}
\author[mainaddress]{J-Ph.~Mols}
\author[mainaddress]{J-P.~Bard}
\author[secondaddress]{J-P.~Dognon}
\author[thirdaddress]{C.~Weinheimer}
\author[fourthaddress]{K.P.~Sch\"{a}fers}
\author[mainaddress]{V.~Sharyy}
\author[mainaddress]{D.~Yvon}

\cortext[correspondingauthor]{Corresponding author}
\address[mainaddress]{CEA, DRF, IRFU, University of Paris-Saclay, F-91191 Gif-sur-Yvette, France}
\address[secondaddress]{CEA, DRF, NIMBE, University of Paris-Saclay, F-91191 Gif-sur-Yvette, France}
\address[thirdaddress]{Institut f\"{u}r Kernphysik, University of M\"{u}nster, Wilhelm-Klemm-Stra{\ss}e 9, D-48149 M\"{u}nster, Germany}
\address[fourthaddress]{European Institute for Molecular Imaging, University of M\"{u}nster, Waldeyerstra{\ss}e 15, D-48149 M\"{u}nster, Germany}

\begin{abstract}
CaLIPSO is an innovative photon detector concept designed for high precision brain PET imaging. For the first time, liquid trimethylbismuth is used as sensitive medium. The detector operates as a time-projection chamber and detects both Cherenkov light and charge signal. Indeed, each 511-keV photon releases a single primary electron that triggers a Cherenkov radiation and ionizes the medium. As trimethylbismuth has never been studied before, we measured its free ion yield defined as the number of electron-ion pairs released by the primary electron. To this end, we developed a low-noise measuring system to determine the weak current induced by a \Co\ source in the liquid with an accuracy better than 5 fA for an electric field up to 7kV/cm. We used tetramethylsilane as benchmark liquid to validate the apparatus and we measured a zero-field free ion yield of 0.53 +/- 0.03 in agreement with literature. However, we found a zero-field free ion yield of 0.083 +/- 0.003 for trimethylbismuth, which is a factor 7 lower than the typical values for similar dielectric liquids. Quantum chemistry computations on heavy atoms tend to demonstrate the high ability of trimethylbismuth to capture electrons, which could explain this weak value. This recombination mechanism marks a new step in understanding charge transport in liquid detectors. Finally, to verify the detectability of individual charge pulses, we developed a charge pulse measurement system which has been successfully validated with TMSi. Measurements with TMBi are ongoing.
\end{abstract}


\end{frontmatter}


\section{Introduction}

\begin{figure}[htb]
\centering 
\includegraphics[width=0.95\textwidth]{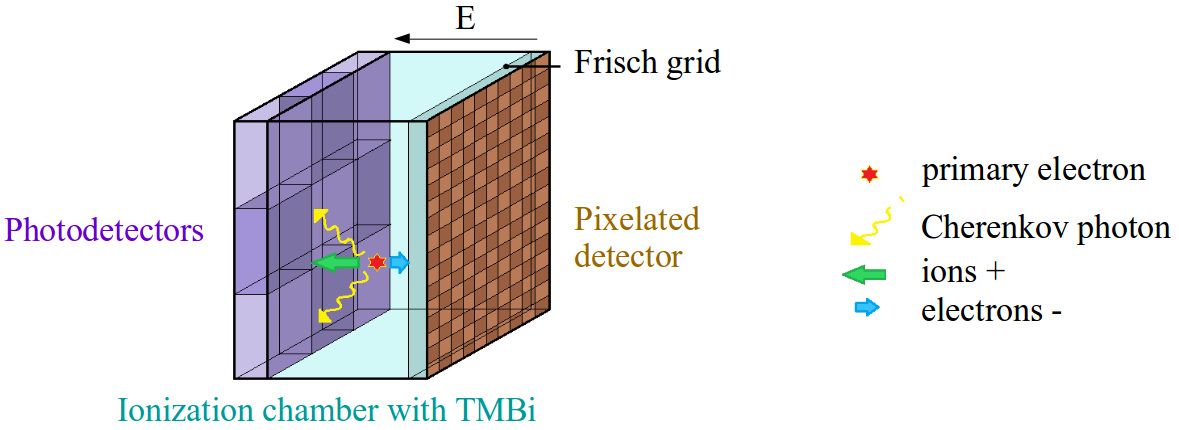}
\caption{Schematic layout of CaLIPSO detector. Liquid trimethyl bismuth allows a double detection: Cherenkov radiation and ionization signal via the charge drift in an electric field.}
\label{fig_calipso}
\end{figure}

CaLIPSO is a photon detector designed for high precision brain PET imaging (Positron Emission Tomography). Liquid trimethylbismuth (TMBi) is used as sensitive medium. Indeed, bismuth is a heavy element ($Z = 83$) and TMBi is a high density liquid (\SI{2.3}{\g\per\cm\cubed}), thereby maximizing the detection efficiency and photoelectric cross section. The 511-keV~photon is thus efficiently converted in a single primary electron.
This electron ionizes the detection medium and releases electron-ion pairs drifted in a strong electric field. This ionization signal allows energy measurement and 2D interaction positioning via a pixelated detector.
The primary electron also produces Cherenkov radiation. The development of photon detection using Cherenkov light gives promising results and is outlined in detail in refs.~\cite{ram16} and \cite{can17}.
CaLIPSO works as a time-projection chamber and detects both ionization and light signals (figure~\ref{fig_calipso}) in order to reconstruct the depth of interaction in the detector. The principle of detection is fully described in ref.~\cite{yvo14}.
The simultaneous double detection leads to promising performances with a precision on 511-keV~photon conversion positioning down to \SI{1}{\mm\cubed} and a coincidence resolution time better than \SI{150}{\pico\s}.

\section{Purification}

During charge pulse measurement, a signal loss can be significant due to the presence of electron-attaching impurities. We designed a purification system based on the protocol written by Engler et al. in ref.~\cite{eng99} for the experiment KASCADE. The liquids in vapor phase are forced through molecular sieves, where polar molecules of small diameters are trapped by 4A and 5A molecular sieves and larger molecules by 13X molecular sieve. TMBi was found to be chemically reactive with 5A and 13X molecular sieves. Only 4A molecular sieve in powder form can thus be used to purify TMBi. This purification system proved effective by gas chromatography-mass spectrometry analysis for a benchmark liquid: tetramethylsilane (TMSi). In the case of TMBi, no impurity was detected with this type of analysis, even before purification.

\section{Free ion yield}

As TMBi has never been studied in particle detection before, we measured the number of electrons released by the primary electron. This is expressed by the free ion yield \Gfi\ defined as the number of electron-ion pairs escaping initial recombination for \SI{100}{\eV} of energy deposited in the liquid. Within the Onsager theory framework~\cite{ons38}, for a moderate electric field $E$, the free ion yield rises linearly with~$E$ as:
\begin{subequations}\label{eq_onsager}
\begin{equation}
\Gfi(E) = \Gfi^0 \ (1+ \alpha E)
\end{equation}
where the slope-intercept ratio $\alpha$ is predicted by:
\begin{equation}\label{eq_alpha}
\alpha = e^3/8\pi\varepsilon(k_BT)^2
\end{equation}
\end{subequations}
where $e$ is the elementary charge, $\varepsilon$ the liquid permittivity, $k_B$ the Boltzmann constant and $T$ the operating temperature.
Onsager treated the problem of the Brownian movement of an electron under the influence of the Coulomb attraction of its parent ion and an additional electric field. This theory can thus be applied when the distance between consecutive ionizations is large when compared to the electron thermalization distance. In TMSi, the average thermalization length measured is \SI{170}{\angstrom}~\cite{hol91}, while the average distance of ionization processes is about \SI{3000}{\angstrom} in hydrocarbon liquids~\cite{eng96}. For TMBi, these distances have not been measured, but we expect the thermalization length and the average distance of ionization processes to be shorter due the higher density of TMBi.

If volume recombination is negligible, the number of electrons released per second is proportional to the current $I$ induced by a radioactive source. Then, the free ion yield \Gfi\ can be expressed as:
\begin{equation}\label{eq_gfi}
\Gfi = I / (e \Delta \epsilon)
\end{equation}
where $e$ is the elementary charge and $\Delta \epsilon$ the energy deposited in the medium per unit of \SI{100}{\eV\per\s}.

\subsection{Methods}

To measure the free ion yield, we engineered a parallel-plate ionization chamber irradiated by a \Co\ source. The chamber is designed with a ceramic body and a gap of \SI{12}{\mm} between the electrodes. The charges produced by radiations are drifted by an electric field and induce a very weak current of the order of a few \SI{10}{\femto\ampere}. Our low-noise measuring device achieves an accuracy better than \SI{5}{\femto\ampere} for a strong electric field up to \SI{7}{\kilo\V\per\cm}. In order to validate our free ion yield measuring device, we first implemented it with liquid TMSi.

\begin{figure}[!htb]
	\centering
	\begin{subfigure}[b]{0.7\textwidth}
		\centering
		\includegraphics[width=\textwidth]{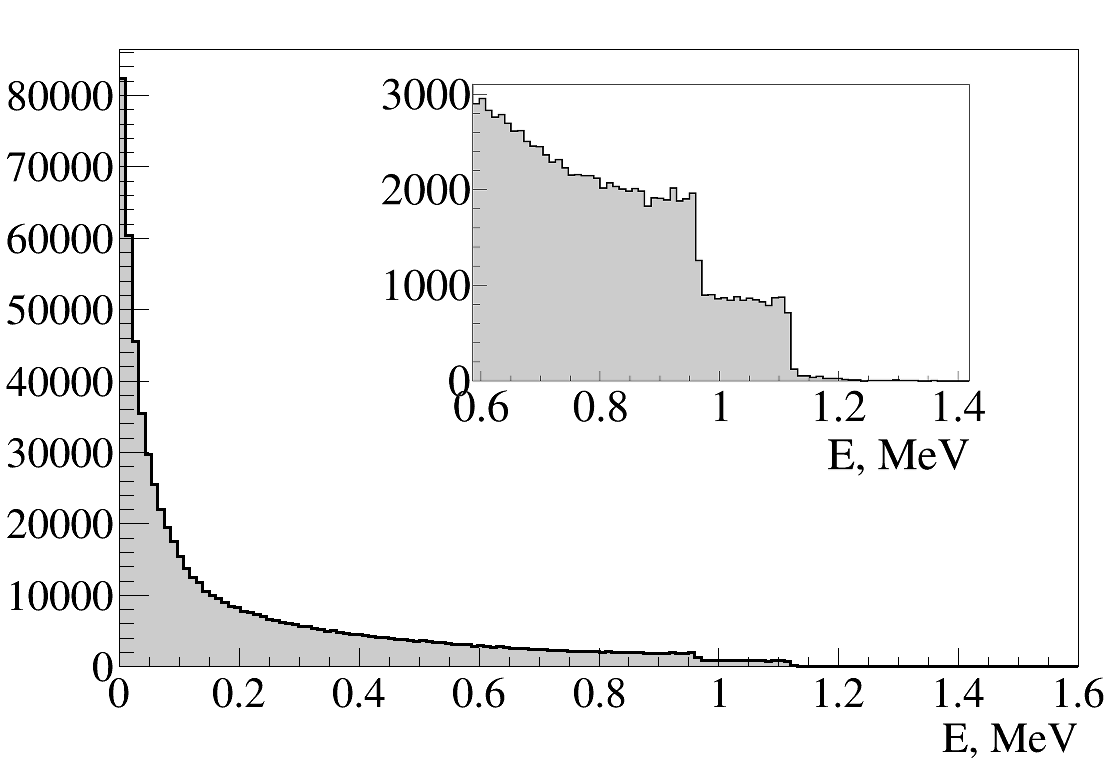}
		\caption{TMSi}\label{fig_edepTMSi}
	\end{subfigure}
	\begin{subfigure}[b]{0.7\textwidth}
		\centering
		\includegraphics[width=\textwidth]{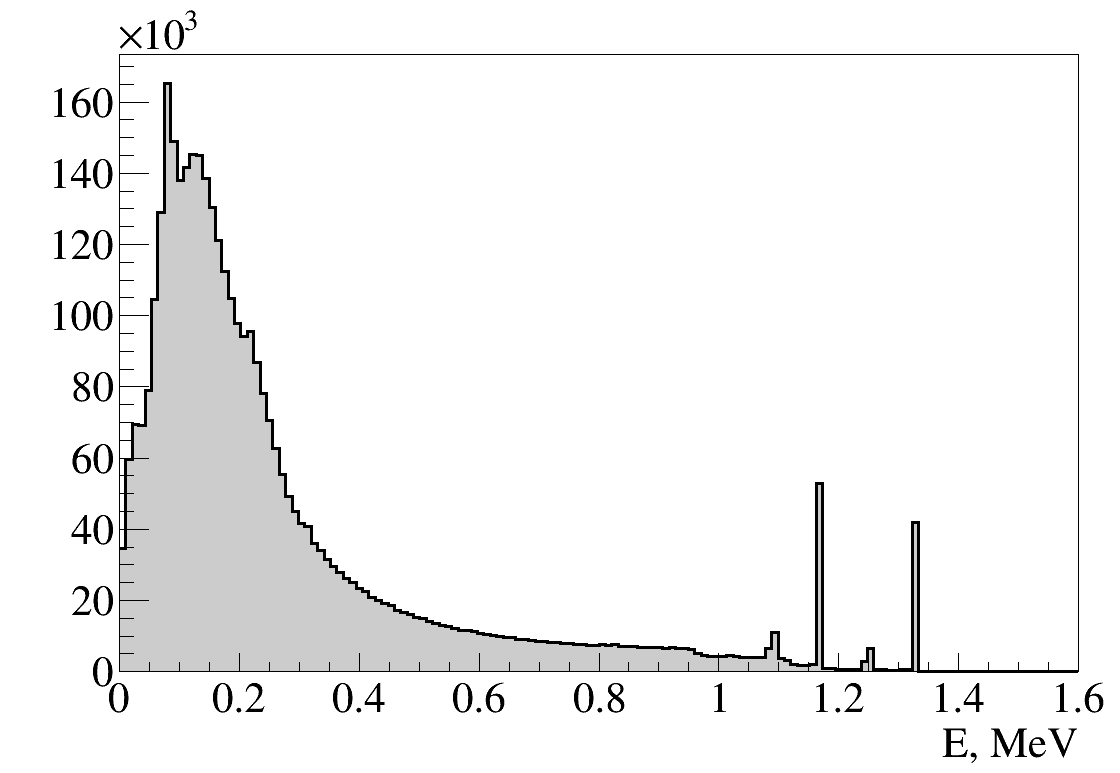}
		\caption{TMBi}\label{fig_edepTMBi}
	\end{subfigure}
\caption{Spectra of energy deposited in the liquids calculated by simulation.}
\label{fig_edep}
\end{figure}

The ionization chamber and its surrounding environment were thoroughly simulated with a \textsc{Gate} simulation~\cite{jan04}. We computed the spectra of energy deposited by the \Co\ source in the liquids to estimate the energy deposited per second (figure~\ref{fig_edep}).

\subsection{Results}

\begin{figure}[htb]
\centering 
\includegraphics[width=0.8\textwidth]{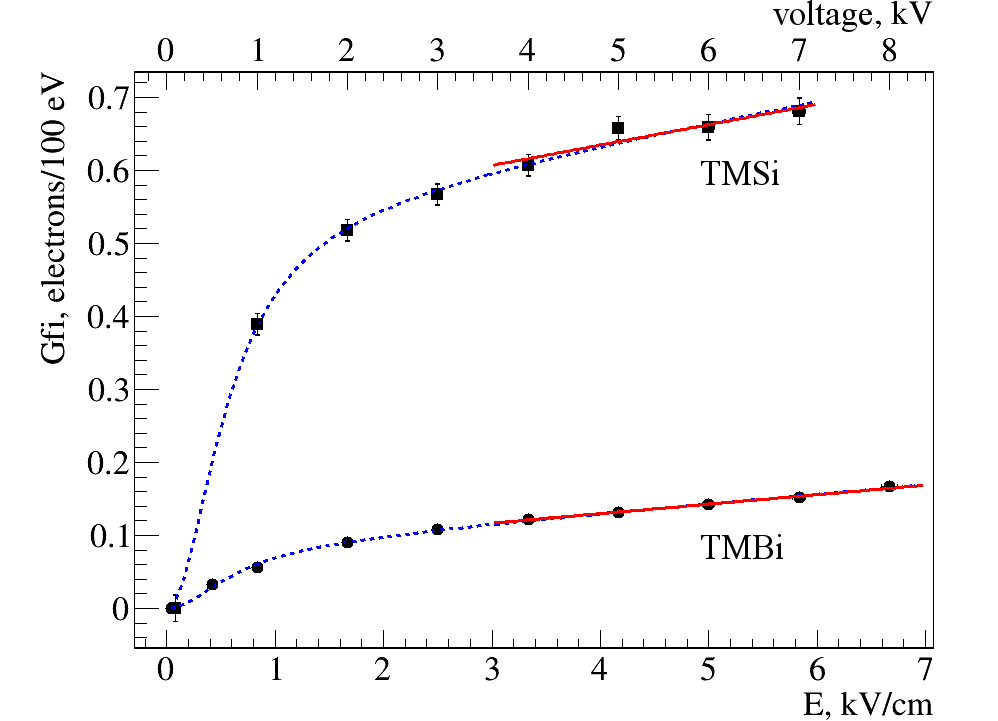}
\caption{Free ion yields of TMSi and TMBi as a function of electric field. The solid red line corresponds to the fit with Onsager equation (eq.~\ref{eq_onsager}).}
\label{fig_Gfi}
\end{figure}

The measurement of the ionization current induced by the \Co\ source in TMSi and TMBi with respect to the applied voltage provides repeatable and reproducible results.
The free ion yield was estimated from eq.~\ref{eq_gfi} and is represented as a function of applied electric field on figure~\ref{fig_Gfi}. We deduced the zero-field free ion yield $\Gfi^0$ and slope-intercept ratio $\alpha$ from Onsager equation (eq.~\ref{eq_onsager}) to which we add a term describing charge collection efficiency as suggested in ref.~\cite{par09}:
\begin{equation}\label{eq_modele}
\Gfi(E) = \Gfi^0 (1+\alpha E) \left[ 1 + \frac{C (1+\alpha E)}{E^2} \right]^{-1}
\end{equation}
where $C$ is a free parameter. This simplified model allows us to describe the non-linear low-field behaviour of free ion yield. The results along with their respective uncertainties are summarized in table~\ref{table_results}. Constant $C$ was found to be $0.29 \pm 0.07$ for TMSi and $0.34 \pm 0.04$ for TMBi.

\begin{table}[htb]
\centering
\caption{Results of zero-field free ion yields and slope-intercept ratios for liquids TMSi and TMBi.}
\label{table_results}
\smallskip
\begin{tabular}{|ll|lll|}
\hline
                   &      & \multicolumn{3}{c|}{$\Gfi^0$}  \\
                   &      & value	& $\pm$stat		& $\pm$syst \\
\hline
Tetramethylsilane & TMSi & 0.53		& $\pm$0.03		& $\pm$0.03 \\
Trimethylbismuth  & TMBi & 0.083	& $\pm$0.003		& $\pm$0.005 \\
\hline
\multicolumn{5}{c}{} \\
\hline
                   &      & \multicolumn{3}{c|}{$\alpha$ (cm$\cdot$kV$^{-1}$)}  \\
                   &      & value	& $\pm$stat		& $\pm$syst	\\
\hline
Tetramethylsilane & TMSi & 0.05	& $\pm$0.02	& $\pm$0.0006 \\
Trimethylbismuth  & TMBi & 0.15	& $\pm$0.02	& $\pm$0.007 \\
\hline
\end{tabular}
\end{table}

\subsection{Discussion}

In the case of TMSi, the measured free ion yield is in agreement with literature. The slope-intercept ratio $\alpha$ predicted by eq.~\ref{eq_alpha} is expected to be \SI{0.0588 \pm 0.0008}{\cm\per\kilo\V} considering a temperature of \SI{20 \pm 2}{\celsius}. Our measurement corroborates this theoretical prediction.
	In the case of TMBi, the free ion yield is a factor 6 lower than the free ion yield of TMSi. Moreover, we expected a slope-intercept ratio $\alpha$ of \SI{0.045 \pm 0.004}{\cm\per\kV}. The measured value is however about 3 times higher than this theoretical prediction. This result shows a breakdown of the Onsager theory to describe recombination in TMBi.

In order to understand these discrepancies, we initiated quantum chemistry calculations. The results, detailed in ref.~\cite{far18}, reflect the global behaviour of TMBi which tends to capture free electrons. Recombination between this anion and the parent cation is then likely to occur due to the high density of TMBi. This additional phenomenon, not included in Onsager theory, is analogous to initial recombination and can explain the weak value of \Gfi\ in our experiment with TMBi.

\section{Charge pulse measurement}

In order to detect individual ionization pulses in TMBi, we consider two different methods. The first one is based on ionizations induced by a \Co\ source and the second one by cosmic muons. First, we measure the lifetime and mobility of electrons in TMSi.

\subsection{Low-noise charge pre-amplifiers}

A charge pre-amplifier has been developed by the Institut f\"{u}r Kernphysik (M\"{u}nster, Germany)~\cite{sch90} and reaches an equivalent noise charge of 79 electrons RMS at ambient temperature with an input capacitance of \SI{0}{\pico\farad} and a rise time (20-80\%) of \SI{218}{\nano\s}. A second pre-amplifier, developed by IRFU/CEA (Saclay, France), is a fast charge amplifier with a measured rise time $T_{20-80}$ of \SI{52}{\nano\s} and an equivalent noise charge of 162 electrons RMS.

\subsection{Charge pulses with \Co\ source}

As a first step, we measure charge pulses induced in the ionization chamber by $\gamma$ photons emitted by the \Co\ source. The signals are amplified by the pre-amplifier developed in M\"{u}nster and shaped by a post-amplifier.

\begin{figure}[htb]
\centering 
\includegraphics[width=0.8\textwidth]{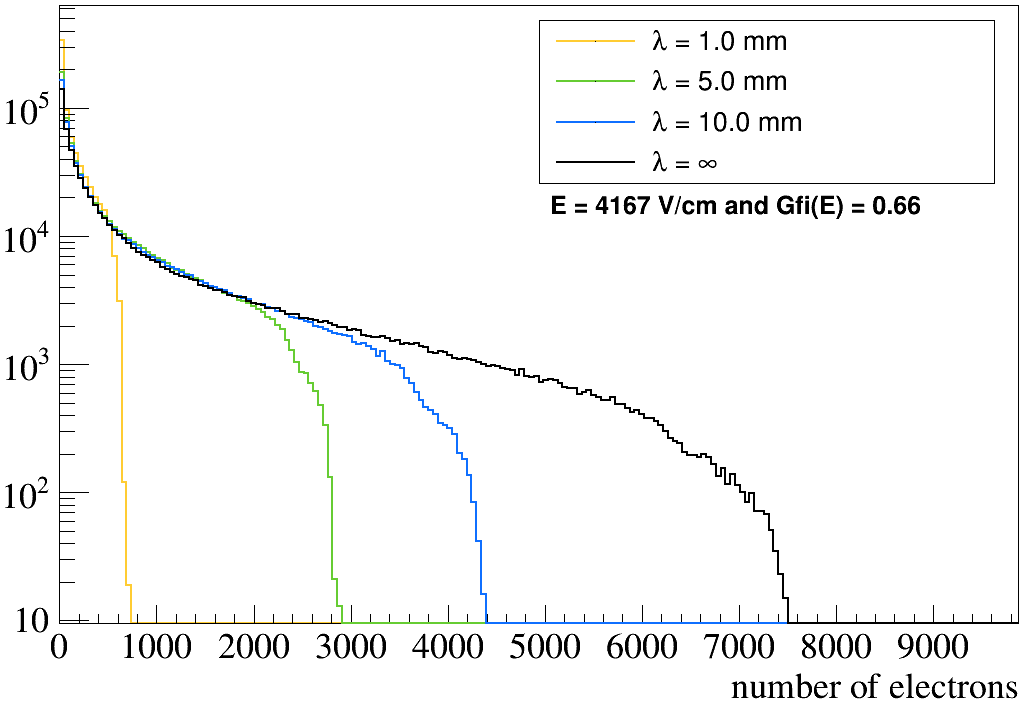}
\caption{Amplitude spectrum of charge pulses induced by the \Co\ source in TMSi as a function of electron mean attachment length $\lambda$.}
\label{fig_chargesource}
\end{figure}

In TMSi, Compton effect is predominant. Therefore, the charge amplitude spectrum presents a Compton edge whose position depends on the free ion yield \Gfi\ and on the electron mean attachment length $\lambda = \tau \cdot \mu E$ where $\tau$ is the electron lifetime, $\mu$ the electron mobility and $E$ the applied electric field.
By comparing the measured Compton edge with the simulation represented on figure~\ref{fig_chargesource}, we were able to estimate an attachment length of \SI{4.7 \pm 0.4}{\mm} (preliminary results) which corresponds to an electron lifetime of \SI{1.05 \pm 0.08}{\micro\s} if the mobility of electrons is \SI{90}{\cm\squared\per\V\per\s}~\cite{sch77}.


\subsection{Charge pulses with cosmic muons}

In the second method, we want to interpret the shape of the rising edge of charge pulses. In order not to degrade this information, we use the pre-amplifier with the fastest rise time, developed in Saclay. The shape of the current induced on the anode by a muon crossing the ionization chamber from one electrode to the other does not depend on the angle of incidence of the muon. Indeed, the induced charge distribution is uniformly distributed along the muon track.
The current induced by the passage of the muon depends on the electron lifetime $\tau$ and mobility $\mu$ as:
\begin{equation}
I(t) = \frac{N e}{t_d} \left( 1-\frac{t}{t_d} \right) \exp \left( -t/\tau \right)  \quad \text{with:} \quad t_d=\frac{d}{\mu E}
\end{equation}
where $N$ is the number of electrons released by the muon, $e$ the elementary charge, $d$ the distance between electrodes and $E$ the applied electric field.

\begin{figure}[htb]
\centering 
\includegraphics[width=\textwidth]{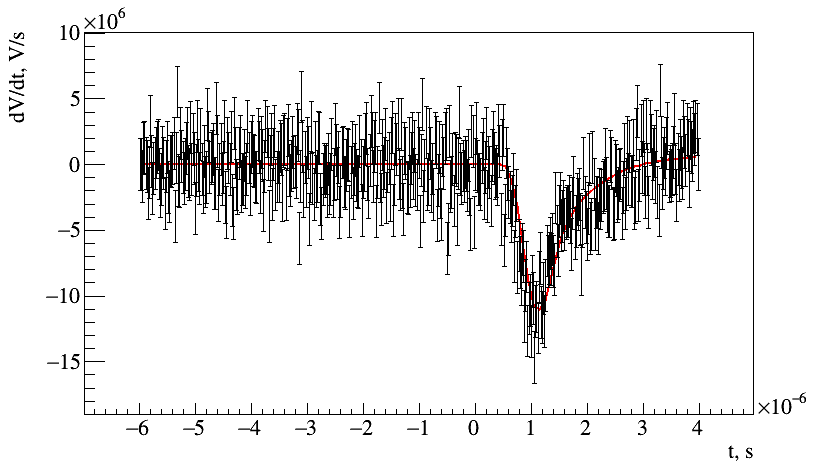}
\caption{Derivative of the measured signal induced by a crossing muon fitted by the convolution of the induced current on the anode and the impulse response, represented in red.}
\label{fig_signal}
\end{figure}

The signal is triggered in coincidence with two muon taggers (plastic scintillators) having a time resolution of the order of a few nanoseconds. We study the impulse response of the charge amplifier connected to the ionization chamber. The measured signal induced by a muon is modelled as the convolution of the induced current on the anode and the impulse response. The electron lifetime value is a free parameter used to reproduce the shape of the measured signal (figure~\ref{fig_signal}). We can deduce: $\tau = \SI{1.1 \pm 0.2}{\micro\s}$ (preliminary results) in consistence with the measurement of charge pulses produced by $\gamma$ photons.

\section{Conclusion}

In this work, we describe the first determination of the free ion yield of liquid TMBi. Thanks to a low-noise measurement of radiation-induced currents, we demonstrated that the zero-field free ion yield of TMBi is a factor 6 lower than of TMSi and the slope-intercept parameter is a factor 3 faster than predicted by the Onsager theory.
To understand this discrepancy, computations in quantum chemistry revealed the ability of TMBi molecules to capture free electrons, which would provide an additional trapping mechanism for ionization electrons near their
parent cations.

To verify the detectability of individual charge pulses induced by a $\gamma$ photon or a charge particle radiation and to measure the electron lifetime in the liquids, we developed a charge pulse measurement system. We validated it successfully with TMSi. Measurements with TMBi are ongoing.

A three-year French-German project has been initiated to demonstrate the feasibility of double detection in TMBi and to develop a prototype of high resolution PET detector (time and spatial).

\section*{Acknowledgements}

This work has been supported by the IDEX Pr\'{e}maturation program 2013, the CEA interdisciplinary program "Technologies pour la Sant\'{e}", the PHC Proscope 2018 Campus france (No. 40556SA) and the joint French-German grant ANR-18-CE92-0012-01 and DFG-SCHA 1447/3-1 and WE 1843/8-1.


\bibliography{References}

\end{document}